# Domain wall conductivity in semiconducting hexagonal ferroelectric TbMnO$_3$ thin films

D. J. Kim[1,2,3,*], J. G. Connell[4], S. S. A. Seo[4], and A. Gruverman[1,†]

[1]*Department of Physics and Astronomy, University of Nebraska-Lincoln, Lincoln, NE 68588, USA*

[2]*Center for Correlated Electron Systems, Institute for Basic Science, Seoul 151-742, Republic of Korea*

[3]*Department of Physics and Astronomy, Seoul National University, Seoul 151-742, Republic of Korea*

[4]*Department of Physics and Astronomy, University of Kentucky, Lexington, KY 40506, USA*

**Although enhanced conductivity at ferroelectric domain boundaries has been found in BiFeO$_3$ films, Pb(Zr,Ti)O$_3$ films, and hexagonal rare-earth manganite single crystals, the mechanism of the domain wall conductivity is still under debate. Using conductive atomic force microscopy, we observe enhanced conductance at the electrically-neutral domain walls in semiconducting hexagonal ferroelectric TbMnO$_3$ thin films where the structure and polarization direction are strongly constrained along the *c*-axis. This result indicates that domain wall conductivity in ferroelectric rare-earth manganites is not limited to charged domain walls. We show that the observed conductivity in the TbMnO$_3$ films is governed by a single conduction mechanism, namely, the back-to-back Schottky diodes model tuned by the segregation of defects.**

---

[*] Author to whom correspondence should be addressed. Electronic mail: muddog@snu.ac.kr

[†] Author to whom correspondence should be addressed. Electronic mail: agruverman2@unl.edu

In the field of oxide and semiconductor electronics, the scientific paradigm has been shifting from bulk to interface[1] as a number of important and interesting phenomena have been discovered at interfaces.[2] A two-dimensional electron gas emerging at interfaces of oxide heterostructures[3] and gradual resistive change in ferroelectric memristive systems[4,5] are representative of interfacial phenomena. The active interfaces in these systems are defined by the fabrication of heterostructures with different materials, namely heterointerfaces. Generally, such heterostructures are very difficult to fabricate because they require coherent growth of each layer and well-defined interfaces.[3]

At the same time, homointerfaces such as ferroic domain walls started to attract attention, because of their intrinsic coherence, nanometer thickness, mobility, and rich physics.[6,7,8] Domain walls are archetypal homointerfaces between regions, which have different orientations of an order parameter within an ordered ferroic material. It has been reported that domain walls exhibit intriguing properties,[7] for example, a finite magnetization in a multiferroic[9] and enhanced/decreased conductivity in ferroelectrics.[10,11,12,13,14] Among them, conductive domain walls in ferroelectrics have great potential to overcome the limitations of conventional electronics by endowing electronic devices with a resistive tunability.

Since the first report of domain walls conduction in multiferroic $BiFeO_3$ thin films,[10] intense studies of this effect in other ferroelectrics have been carried out.[11,12,13,14,15,16,17] The recent discovery of enhanced/reduced conductance at domain walls in $Pb(Zr,Ti)O_3$ films,[18] $LiNbO_3$ single crystals,[19] and hexagonal rare-earth manganite ($h$-$Re$MnO$_3$) single crystals[15,16,17] suggests that the domain wall conductivity is universal in ferroelectrics. In particular, the domain wall conductivity in *improper* ferroelectric $h$-$Re$MnO$_3$ should be more intriguing due to the presence of energetically unfavorable charged domain walls stabilized by the interlocking of structural phase boundaries and domain walls, and the presence of topological defects and vortices.[20] The recently reported conductivity at domain walls in $h$-HoMnO$_3$ and $h$-ErMnO$_3$ can be understood as a combined consequence of changes in the band structure at the domain walls and accumulation of the charge carriers compensating the charged domain walls.[16,17] However, studies of domain wall conductivity in $h$-$Re$MnO$_3$ have thus far been focused on single crystals, while there has been no study on thin films. Here, we report enhanced domain wall conductivity in room-temperature ferroelectric $h$-TbMnO$_3$ thin films[21] epitaxially stabilized on Pt/Al$_2$O$_3$ substrates.[22]

Epitaxial $h$-TbMnO$_3$ thin films were grown on Pt(111)/Al$_2$O$_3$(0001) substrates using pulsed laser deposition (PLD) system equipped with *in situ* reflection high-energy electron diffraction (RHEED) and optical spectroscopic ellipsometry (SE).[23] Prior to deposition of the $h$-TbMnO$_3$ film, a Pt bottom layer was deposited on an (0001) Al$_2$O$_3$ single crystal substrate at 600 °C using rf-sputtering. The optimal growth conditions of $h$-TbMnO$_3$ thin films in PLD were found to be the following: a substrate temperature of 800 °C, an oxygen partial pressure of 30 mTorr, and a KrF excimer ($\lambda$ = 248 nm) laser fluence of 1.2 J/cm$^2$. The coherent growth of hexagonal TbMnO$_3$ film was confirmed by monitoring with RHEED and SE during the deposition. The thickness of the deposited $h$-TbMnO$_3$ film was 20 unit cells, equivalent to ~23 nm. X-ray diffraction scans revealed the film to be hexagonal and epitaxially grown.[21] Atomic force microscopy (MFD-3D, Asylum Research) was used to acquire local conductance maps and current-voltage (I-V) curves at different fixed locations with an electrically biased Pt-coated tip (PPP-EFM, Nanosensors). Bias for writing a resistance state or reading a current was applied to the tip, and the bottom electrode was grounded.

In our previous work,[21] room-temperature ferroelectricity and polarization-dependent resistive switching in $h$-TbMnO$_3$ thin films have been reported. The subsequent observations of a local conductance with different read-bias revealed the enhanced domain wall conductivity of ferroelectric $h$-TbMnO$_3$ thin films. In Fig. 1(a), the difference in conductance between negatively (OFF; −6 V$_{write}$) and positively (ON; 6 V$_{write}$) poled areas at the -0.5 V read-bias of the tip, is barely visible. There are mesas of a low resistance, which are not identified but believed to be structural/vacancy defects. With an increased read-bias (in the range of -1.0 ~ -1.5 V), conduction of domain walls becomes visible (Figs. 1(b) & 1(c)). Only the conduction at the domain walls between the OFF domain and the as-grown (unpoled) region is visible, while the conduction at the domain walls between the OFF and the ON domains is not, because the strong conduction of the ON domain conceals that of the domain walls. In other words, the domain wall conduction around the ON domain cannot be distinguished from the conduction of the ON domain itself. With a higher read-bias (-2.0 V, Fig. 1(d)), the conduction of the domain walls is not seen because it is not distinguishable from the increased conduction of the surrounding unpoled region. The surrounding unpoled region is believed to be associated with a polydomain structure, which a PFM tip does not allow one to visualize due to the small domain size. There was no modification of the topography and the current map by repeated scanning with the biased tip, which indicates that the conductance change is not due to surface damage by the tip and that the reading voltage does not affect the conductance state of the scanned area.

The domain wall conduction, as well as the conduction within the ON domain and the unpoled region, is not uniform. Figure 2(a) shows a histogram analysis of the current distribution within the ON and OFF domains, the unpoled region and the domain wall, acquired from Fig. 1(c). Each distribution is described well by a lognormal function, except for the OFF region, which was too narrow to fit. A lognormal distribution of current implies a Gaussian distribution of the conduction barrier properties, namely effective barrier thickness and height.[24,25] Given that the surface roughness does not have sufficient spatial variation to produce a broad lognormal distribution (the root-mean-square roughness of the surface from which the data of Fig. 2 have been taken is less than 0.5 nm, while the film thickness is ~23 nm), it is expected that the conduction through the film would be governed mainly by the effective conduction barrier height, not by the barrier thickness. From the distribution and the averaged current values, as shown in Fig. 2(b), we found that the states' resistances $R_{OFF} > R_{unpoled} > R_{OnWall} > R_{ON}$.

Figures 3(a)-(d) show I-V curves measured at many locations randomly chosen in the ON and OFF domains, the unpoled region, and the domain wall. Clearly, all of the observed conduction through the film is not a quantum mechanical tunneling since the film is too thick (~23 nm). Given that $h$-TbMnO$_3$ is a narrow band gap semiconductor with a 1.4 eV band gap,[26] a model of back-to-back Schottky barriers can be invoked to explain the observed I-V curves.[21] This model has two fitting parameters, the Schottky barrier height and the ideality factor of the interfaces. The ideality factor reflects non-ideal effects such as interface states, image-forces, etc. Unfortunately, due to a lack of physical information such as the effective contact area between the tip and the film surface and due to the strong spatial inhomogeneity of the I-V curves, we could not extract meaningful barrier height values from I-V curve fitting with this model. Interestingly, as reported earlier[21] and as seen in Fig. 3(e), all I-V curves once normalized by their values at 2.5 V can be well described by a single universal fitting curve of the back-to-back Schottky barriers model.[27] This means that the ideality factor for all the conduction can be specified as a universal value while one can obtain only qualitative information of the Schottky barrier height.[21] From the averaged current values in Fig. 2(b), the differences between the barrier heights are roughly 0.17 eV (ON vs. OFF), 0.15 eV (On Wall vs. OFF), and 0.13 eV (Unpoled vs. OFF). In conclusion, all the conduction in the ON and OFF domains, the unpoled region, and at the domain wall has the universal

ideality factor of 1.09 and the barrier height is the sole parameter governing the resistance state.

The exact mechanism of conductivity at ferroelectric domain walls is still under debate.[7,8,18,19,20] So far, several possible mechanisms have been proposed: (1) an electrostatic potential step due to the discontinuity of in-plane polarization at the domain wall;[10,11] (2) a lowering of the band gap due to structural changes across the domain wall;[10] and (3) a strain-associated segregation of oxygen vacancies at the domain wall, resulting in a lowering of the Schottky barrier.[14] The first explanation cannot be applied in our case because the polarization in our sample has no in-plane component and, therefore, the polarization component normal to the wall should be zero. The neutral 180° domain walls in an epitaxial Pb(Zr,Ti)$O_3$ film may have some charged portion by forming a flux-closure domain formation,[18] given the direct observation of continuous polarization rotation at the domain wall with transmission electron microscopy (TEM).[28] However, in our $h$-TbMnO$_3$ films, as the domain walls and the structural antiphase boundaries in $h$-$Re$MnO$_3$ are interlocked tightly,[20,29] a continuous rotation of polarization producing a charged domain wall is not likely. It would be of interest to investigate the local domain configurations at the domain walls in this material by high-resolution TEM.

One can assume that a structural change at the structural antiphase boundary (interlocked with a domain wall) may reduce the band gap,[7,10,15] providing a conduction path. However, recent first-principle calculations showed that the local band gap does not change at the neutral domain walls in $h$-YMnO$_3$,[29] so the possibility of band gap modification at the structural antiphase boundary can be ruled out.

The accumulation of oxygen vacancies at the domain walls,[13,14] which results in a reduced Schottky barrier and a conduction path,[30] is thus the most probable mechanism underlying the domain wall conduction in the $h$-TbMnO$_3$ film. The oxygen vacancies can be considered as mobile positive charges, able to migrate to thermodynamically stable positions under applied electric field. When a domain is written, a biased tip sweeps away oxygen vacancies inside the film, which are accumulated at the structural antiphase boundary. This scenario is consistent with the random distribution of effective conduction barriers at the domain walls. Moreover, the fact that the domain wall conduction can evidently be tuned by the density of oxygen vacancies in an $h$-YMnO$_3$ single crystal[15] supports the interpretation that oxygen vacancies play the most important role in the domain wall conduction of $h$-$Re$MnO$_3$.

In summary, we have investigated the domain wall conductivity in ferroelectric hexagonal TbMnO$_3$ thin films using conductive atomic force microscopy. Besides charged domain wall conductivity in $h$-$Re$MnO$_3$ films and single crystals, neutral domain walls of hexagonal TbMnO$_3$ film show enhanced conductivity, indicating the domain wall conductivity to be a universal property. Although further investigations are required to confirm the details of the domain wall conductivity of the hexagonal TbMnO$_3$ film, a pure electrostatic/electronic effect in association with defect chemistry based on the back-to-back Schottky diodes structure should be considered.

**Acknowledgements**

Research at the University of Nebraska-Lincoln was supported by the US Department of Energy, Materials Sciences Division, under Award No. DE-SC0004876 (electrical characterization) and the National Science Foundation (NSF) through the Nebraska Materials Research Science and Engineering Center (MRSEC) under Grant No. DMR-1420645 (modeling). The work (sample preparation) at the University of Kentucky was supported by the NSF through Grant No. EPS-0814194 (the Center for Advanced Materials) and by the Kentucky Science and Engineering Foundation with the Kentucky Science and Technology Corporation through Grant Agreement No. KSEF-148-502-14-328. D. J. K. was supported by IBS-R009-G1.

**Figure captions**

**FIG 1.** Local current images of hexagonal TbMnO$_3$ film obtained with (a) -0.5 V, (b) -1.0 V, (c) -1.5 V, and (d) -2.0 V biased tip. Negatively poled areas (upper) show a high resistance state while positively poled areas (lower) show a low resistance state. The scan size is 6 × 6 μm$^2$.

**FIG 2.** (a) (Left) Distribution of current acquired on high (OFF) and low (ON) resistance areas, on a domain wall, and on an unpoled as-grown area with −1.5 V read bias. (Right) Every distribution is well described by a lognormal function except for OFF due to that distribution being too narrow. (b) Averaged values of current and center values of fitted lognormal functions acquired on the OFF and ON areas, on the domain wall, and on the as-grown unpoled region, respectively. Error bars indicate standard deviation of current and width of the distributions.

**FIG 3.** I-V curves measured at many spots in (a) the ON domain, (b) at the domain wall, (c) in the unpoled region, and (d) in the OFF domains of a hexagonal TbMnO$_3$ film. (e) I-V curves normalized by their values at 2.5 V of (a) ~ (d). The red dashed line is a fitting result using the back-to-back Schottky barrier model with an ideality factor of 1.09 and an arbitrary barrier height. The normalized I-V curves of OFF, unpoled, and ON and the fitting curve are reproduced from Ref. 21.

**Figure 1**

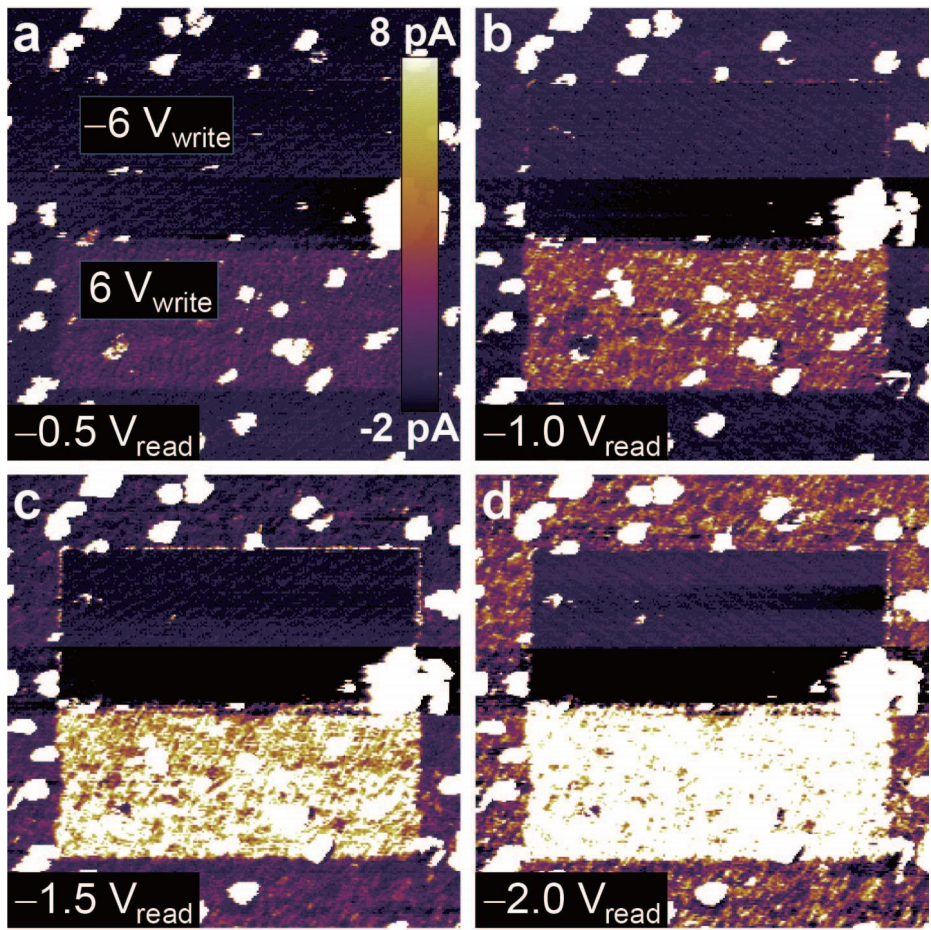

**Figure 2**

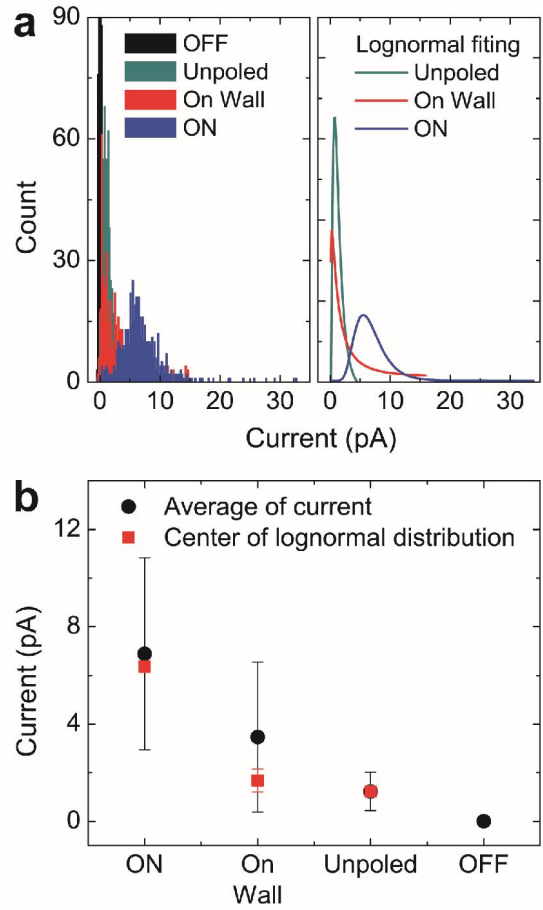

**Figure 3**

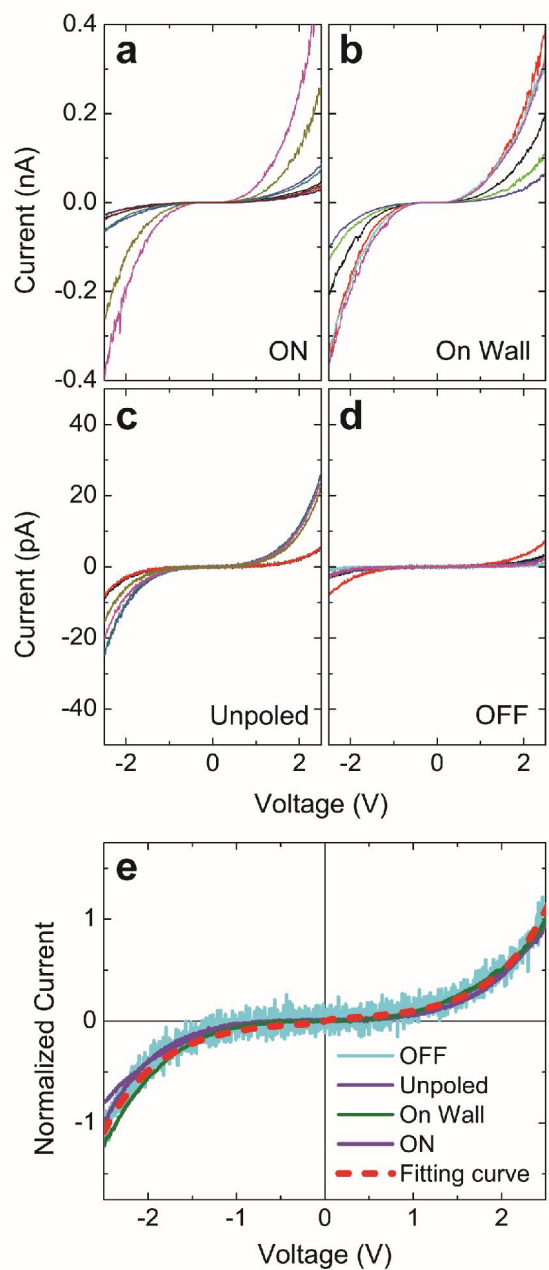
12